\documentclass[aip,jcp,preprint,groupedaddress]{revtex4-1}
 
\usepackage{graphicx}
\usepackage{dcolumn}
\usepackage{bm}
\usepackage{amsmath}
\usepackage{textcomp}
\usepackage{longtable}
\usepackage{url}

\begin{document}

\title{A first-principles study of structural and elastic properties of bulk SrRuO$_3$}

\author{\v{S}. Masys }
\email[Author to whom correspondence should be addressed. Electronic mail: ]{Sarunas.Masys@tfai.vu.lt}

\author{V. Jonauskas}

\affiliation{Institute of Theoretical Physics and Astronomy, Vilnius University, A. Go\v{s}tauto Street 12, LT-01108 Vilnius, Lithuania}

\date{\today}

\begin{abstract}
We present a first-principles investigation of structural and elastic properties of experimentally observed phases of bulk SrRuO$_3$ - namely orthorhombic, tetragonal, and cubic - by applying density functional theory (DFT) approximations. At first, we focus our attention on the accuracy of calculated lattice constants in order to find out DFT approaches that best represent the crystalline structure of SrRuO$_3$, since many important physical quantities crucially depend on change in volume. Next, we evaluate single-crystal elastic constants, mechanical stability, and macroscopic elastic parameters trying to at least partially compensate for the existing lack of information about these fundamental features of SrRuO$_3$. Finally, we analyze the anomalous behavior of low-temperature orthorhombic phase under $C_{44}$ related shear deformation. It turns out that at critical strain values the system exhibits a distinct deviation from the initial behavior which results in an isosymmetric phase transition. Moreover, under $C_{44}$ related shear deformation tetragonal SrRuO$_3$ becomes mechanically unstable raising an open question of what makes it experimentally observable at high temperatures.     
 
\end{abstract}

\maketitle

\section{Introduction}

Strontium ruthenate SrRuO$_3$ is a perovskite-type metallic oxide exhibiting ferromagnetic properties below Curie temperature of 160 K (Refs. \onlinecite{mazin_1997_1,kim_2005_2,siemons_2007_3}). The combination of high resistance to corrosion \cite{herranz_2005_4}, excellent electrical conductivity, strong chemical stability, and easy epitaxial growth on various functional transition metal oxides \cite{eom_1992_5,eom_1993_6,wu_1993_7,he_2004_8} has attracted noticeable attention of scientific community. SrRuO$_3$ is a promising material for electrodes and junctions of perovskite-based devices, including ferroelectric heterostructures and non-volatile ferroelectric random access memories \cite{shin_2005_9}, since its good thermal conductivity and surface stability help to improve retention, fatigue resistance, and imprint \cite{palai_2009_10}.

At room temperature bulk SrRuO$_3$ crystallizes in an orthorhombic (space group $Pbnm$) GdFeO$_3$-type perovskite structure, in which the RuO$_6$ octahedra are tilted \cite{jones_1989_11}. As the temperature increases up to 950 K, this compound undergoes a series of phase transformations, in detail described in Refs. \onlinecite{kennedy_1998_12,chakoumakos_1998_13,yamanaka_2004_14}. An orthorhombic structure of SrRuO$_3$ consecutively transforms to tetragonal $I4/mcm$ (stable in the temperature range of 820 to 950 K) and then to cubic $Pm\bar{3}m$ symmetry. Although by using high-resolution synchrotron diffraction and fine temperature intervals Kennedy $et$ $al$. \cite{kennedy_2002_15} have revealed the presence of intermediate $Imma$ phase between 685 and 825 K, due to the lack of the more thorough experimental analysis of the corresponding SrRuO$_3$ structure we did not include it in the present work. Concerning the recent low-temperature studies, Bushmeleva $et$ $al$. \cite{bushmeleva_2006_16} have reported the neutron diffraction results of an orthorhombic phase at 1.5 K. It is worthwhile to mention that below 160 K SrRuO$_3$ crystalline structure undergoes only tiny changes due to the Invar effect initially discovered by Kiyama $et$ $al$. \cite{kiyama_1996_17}

The main goal of our study is to systematically investigate the geometric structure and elastic properties of experimentally observed phases of SrRuO$_3$ by applying the most recent generalized-gradient approximations (GGAs) for exchange-correlation energy of Kohn-Sham density functional theory (DFT), namely PBEsol (Ref. \onlinecite{pbesol_18}), SOGGA (Ref. \onlinecite{sogga_19}), and WC (Ref. \onlinecite{wc_20}). Since many inherent material properties including phonon frequencies, elastic constants, and ferromagnetism are critically dependent on change in volume \cite{csonka_2009_21,haas_2010_22}, primarily we focus our attention on the accuracy of the calculated lattice constants. It has been previously noticed that for various perovskite crystals local density approximation (LDA) systematically underestimates the experimental lattice constants on average by 2\%, whereas standard GGA overestimates them by the same magnitude \cite{nishimatsu_2010_23}. Although in general these discrepancies seem to be acceptable, absolute relative errors of the ``good" theoretical values should not exceed 0.5\% (Ref. \onlinecite{haas_2009_24}). But despite the aforementioned trend, we have incorporated LDA (Refs. \onlinecite{lda_25, pz_26}) and PBE (Ref. \onlinecite{pbe_27}) calculations in the present work to make our analysis more complete. However, we did not take into consideration hybrid functionals due to the metallic character of SrRuO$_3$ and arguments for its weakly correlated nature \cite{etz_2012_28,masys_2013_29}. It is well known that hybrids are very fruitful for systems possessing band gaps and in some cases the inclusion of the exact Hartree-Fock (HF) exchange energy may serve as a reasonable alternative for the description of strongly correlated electrons \cite{masys_2010_30,gou_2011_31}. These issues, though, are unlikely to be relevant for bulk SrRuO$_3$. 

The crystalline structure calculations of SrRuO$_3$ available so far were done at its orthorhombic and cubic phases using the LDA framework (see, e.g., Refs. \onlinecite{zayak_2006_32,albina_2007_33,rondinelli_2008_34,hadipour_2011_35}). A very recent work of Garc\'{i}a-Fern\'{a}ndez $et$ $al$. \cite{garcia_2012_36} also includes PBE, PBEsol, and WC functionals together with several combinations of their hybrid schemes. It is somehow surprising that in the literature scarcely any first-principles calculations of tetragonal SrRuO$_3$ have been reported to date. Even more intriguingly, one can hardly find any theoretical as well as experimental investigation on the elastic properties of SrRuO$_3$, except for several polycrystalline measurements for the orthorhombic phase \cite{yamanaka_2004_14,dabrowski_2005_37,hamlin_2007_38}. In order to at least partially remedy this gap, we have performed an extensive theoretical analysis of experimentally determined phases of SrRuO$_3$ starting from the accuracy of the evaluated lattice constants, then concentrating on elastic properties and mechanical stability, and finally providing some insights into anomalous behavior caused by shear deformation. The obtained results are accordingly presented and discussed in Sec. \ref{sec4} A, B, and C.

\section{Theoretical background}
\subsection{DFT approaches}

In the LDA framework, the exchange energy $E_{X}$ has the form
\begin{equation}
E_{X}^{LDA}[n]=\int n\varepsilon_{X}^{LDA}(n)d^{3}r,
\label{eq:equ1}
\end{equation}
where $n$ denotes the electron density and $\varepsilon_{X}^{LDA}(n)=-(3/4)(3/\pi)^{1/3}n^{1/3}$ is the exchange energy density per particle for a uniform electron gas. The GGA form for the exchange energy is simply
\begin{equation}
E_{X}^{GGA}[n]=\int n\varepsilon_{X}^{LDA}(n)F_{X}(s)d^{3}r,
\label{eq:equ2}
\end{equation}
in which $s=\vert\nabla n\vert/(2k_{F}n)$ (with Fermi wave vector $k_{F}=(3\pi^{2}n)^{1/3}$) is the dimensionless reduced gradient and $F_{X}(s)$ is the exchange enhancement factor. Any GGA that reproduces the uniform gas limit can be expressed as \cite{pbesol_18}
\begin{equation}
F_{X}(s)=1+\mu^{GE}s^{2}+\ldots (s\rightarrow 0),
\label{eq:equ3}
\end{equation}
and accordingly
\begin{equation}
E_{X}^{GGA}[n]=\int n\varepsilon_{X}^{LDA}(n)\lbrace 1+\mu^{GE}s^{2}+\ldots \rbrace d^{3}r=E_{X}^{LDA}[n]+\int n\varepsilon_{X}^{LDA}(n)\lbrace \mu^{GE}s^{2}+\ldots \rbrace d^{3}r,
\label{eq:equ4}
\end{equation}
where the gradient expansion (GE) that is precise for slowly-varying electron gases has \cite{antono_1985_39}
\begin{equation}
\mu^{GE}=10/81\approx0.1235 .
\label{eq:equ5}
\end{equation}
The PBE GGA is nowadays considered as a standard functional for solid-state calculations \cite{haas_tran_2009_40}. Although this GGA belongs to the class of parameter-free functionals, it still contains some arbitrary choices, e.g., the analytical form of the enhancement factor or the constraints that have to be satisfied. It has been shown \cite{pbesol_18,perdew_2006_41} that in order to obtain the accurate exchange energy for free neutral atoms, any GGA must have $\mu\approx2\mu^{GE}$. For the PBE functional, $\mu$ is set to 0.2195 from slightly different requirement which is based on the reproduction of the LDA jellium response. Here, $F_{X}(s)$ has the form
\begin{equation}
F_{X}^{PBE}(s)=1+\kappa\left(1-\frac{1}{1+\frac{\mu s^{2}}{\kappa}}\right).
\label{eq:equ6}
\end{equation}
The parameter $\kappa$, which controls the behavior at $s \rightarrow \infty$, is set to 0.804 according to the relation $\kappa=\lambda_{LO}/2^{1/3}-1$ to ensure the Lieb-Oxford (LO) bound \cite{lieb_1981_42}, which is an upper limit on the ratio of the exact exchange-correlation energy to the value of the LDA approximation of the exchange energy $(E_{X}[n]\geq E_{XC}[n]\geq \lambda_{LO}E_{X}^{LDA}[n]$ with $\lambda_{LO}=2.273)$. At the limit of a slowly-varying high density, the GE of the correlation energy of a GGA can be written as
\begin{equation}
E_{C}^{GGA}[n]=\int n\lbrace \varepsilon_{C}^{LDA}(n)+\beta^{GE}t^{2}+\ldots \rbrace d^{3}r=E_{C}^{LDA}[n]+\int n\lbrace \beta^{GE}t^{2}+\ldots \rbrace d^{3}r,
\label{eq:equ7}
\end{equation}
where $\varepsilon_{C}^{LDA}(n)$ is the correlation energy per particle of the uniform electron gas, $\beta^{GE}$ is a coefficient set to 0.0667 (Ref. \onlinecite{ma_1968_43}), and $t=\vert\nabla n\vert/(2k_{TF}n)$ (with Thomas-Fermi screening wave vector $k_{TF}=\sqrt{4k_{F}/\pi}$) denotes an appropriate reduced density gradient for correlation. In the PBE correlation functional, the value of $\beta^{GE}$ is retained, whereas in PBEsol, it is chosen to be 0.046 in order to reproduce the accurate exchange-correlation energies for a jellium surface obtained at the meta-GGA TPSS (Ref. \onlinecite{tao_2003_44}) level. In the PBEsol exchange functional, the value of $\mu$, which determines the behavior for $s\rightarrow 0$, is restored back to $\mu^{GE}=10/81$, since it has been argued \cite{pbesol_18} that $\mu\approx2\mu^{GE}$ is harmful for many condensed matter applications. This choice allows to recover the second-order GE, but on the other hand, it means that PBEsol no longer satisfies the LDA jellium response, because $\mu\neq \pi^{2}\beta/3$. Thus, as $s\rightarrow 0$, there is no complete cancellation between beyond-LDA exchange and correlation contributions. While constructing the PBE functional, this sort of cancellation was believed to be more accurate than the lower-order gradient expansion for small $s$. Although PBE works equally well for finite ant infinite systems \cite{haas_tran_2009_40}, PBEsol outperforms it in many crystal structure calculations. Nevertheless, this benefit is accompanied by a worsening of the thermochemical properties \cite{sogga_19}. It is evident that due to pretty simple mathematical form of GGA one has to choose between improved atomization energies of molecules or improved lattice parameters of solids \cite{haas_2010_22}.    

The SOGGA exchange functional, used in combination with the PBE correlation functional, completely restores the GE to the second order for both exchange and correlation \cite{sogga_19}. The analytical form of the SOGGA exchange enhancement factor is expressed as an average of the PBE and RPBE (Ref. \onlinecite{hammer_1999_45}) exchange functionals
\begin{equation}
F_{X}^{SOGGA}(s)=1+\kappa\left(1-\frac{1}{2}\cdot \frac{1}{1+\frac{\mu s^{2}}{\kappa}}-\frac{1}{2}\cdot e^{-\mu s^{2}/\kappa}\right),
\label{eq:equ8}
\end{equation} 
in which $\mu=\mu^{GE}=10/81$. The parameter $\kappa$ is set to 0.552 in order to satisfy a tighter LO bound $(E_{X}[n]\geq E_{XC}[n]\geq \lambda_{tLO}E_{X}^{LDA}[n]$ with $\lambda_{tLO}=1.9555)$. 

The WC exchange enhancement factor is given by \cite{wc_20}
\begin{equation}
F_{X}^{WC}(s)=1+\kappa\left(1-\frac{1}{1+\frac{x(s)}{\kappa}}\right),
\label{eq:equ9}
\end{equation} 
where
\begin{equation}
x(s)=\frac{10}{81}s^{2}+\left(\mu - \frac{10}{81} \right)s^{2}e^{-s^{2}}+ln(1+cs^{4}).
\label{eq:equ10}
\end{equation} 
Parameters $\kappa$ and $\mu$ have the same values as in PBE and $c=0.0079325$ is set to recover the fourth order parameters of the fourth order GE of the exchange functional for small $s$ (unfortunately, incorrectly \cite{haas_tran_2009_40}). On the whole, the analysis of a large set of solids \cite{haas_2009_24} shows that concerning the lattice constants PBEsol, SOGGA, and WC perform quite similarly, and in most cases demonstrate an explicit improvement over LDA and PBE. 
 
We would like to remark that in this paper PBE, SOGGA, and WC exchange functionals are used with the correlation part of PBE, whereas the PBEsol exchange functional is employed with the correlation part of PBEsol and PBE. A separate notation PBEsol\textsuperscript{PBE} is introduced for the latter combination.

\subsection{Elastic constants}

Under a linear elastic deformation, solid bodies are described by Hooke's law. In the tensorial form, it can be expressed as
\begin{equation} 
\sigma_{ij}=\displaystyle\sum\limits_{k,l=1}^{3}C_{ijkl}\varepsilon_{kl}, 
\label{eq:equ11}
\end{equation}
where $i$, $j$, $k$, and $l$ are the indices running from 1 to 3, $\sigma_{ij}$ is the stress, $\varepsilon_{kl}$ is the strain, and $C_{ijkl}$ is a fourth-rank stiffness or elastic constants tensor having 81 components. In general, elastic constants describe the material's resistance against an externally applied strain. The symmetry relation $C_{ijkl}=C_{jikl}=C_{ijlk}=C_{klij}$ reduces the number of independent components to 21, which in turn can be further reduced if the material under consideration possesses its own symmetry. According to Voigt notation ($11 \rightarrow 1$, $22 \rightarrow 2$, $33 \rightarrow 3$, $23=32 \rightarrow 4$, $13=31 \rightarrow 5$, $12=21 \rightarrow 6$), the $C_{ijkl}$ components can be arranged in a symmetric $6 \times 6$ matrix. Then, (\ref{eq:equ11}) can be written as \cite{grimvall_46}
\begin{equation} 
\sigma_{\alpha}=\displaystyle\sum\limits_{\beta=1}^{6}C_{\alpha\beta}\varepsilon_{\beta}, 
\label{eq:equ12}
\end{equation} 
where 
\begin{equation} 
\sigma_{\alpha}=\sigma_{ij},
\label{eq:equ13} 
\end{equation} 
\begin{equation} 
\varepsilon_{\beta}=\varepsilon_{kl} \text{ if } \beta=1, 2 \text{ or } 3,
\label{eq:equ14}
\end{equation} 
\begin{equation} 
\varepsilon_{\beta}=2\varepsilon_{kl} \text{ if } \beta=4, 5 \text{ or } 6.
\label{eq:equ15}
\end{equation} 
The relation (\ref{eq:equ12}) for the stresses expressed in the strains can be inverted to give strains in terms of the stresses
\begin{equation} 
\varepsilon_{\alpha}=\displaystyle\sum\limits_{\beta=1}^{6}S_{\alpha\beta}\sigma_{\beta}, 
\label{eq:equ16}
\end{equation} 
in which the compliance matrix $S_{\alpha\beta}$ is inverse to the stiffness matrix $C_{\alpha\beta}$.

The total energy of the distorted crystal's unit cell can be expressed through a Taylor series in terms of the strains
\begin{equation}
E(\varepsilon)=E_{0}+\displaystyle\sum\limits_{\alpha=1}^{6}\frac{\partial E(\varepsilon)}{\partial\varepsilon_{\alpha}}\bigg|_{\varepsilon=0}\varepsilon_{\alpha}+\frac{1}{2}\displaystyle\sum\limits_{\alpha,\beta=1}^{6}\frac{\partial^{2}E(\varepsilon)}{\partial\varepsilon_{\alpha}\partial\varepsilon_{\beta}}\bigg|_{\varepsilon=0}\varepsilon_{\alpha}\varepsilon_{\beta}+\dots,
\label{eq:equ17}
\end{equation}
where $E_{0}=E(0)$ refers to the equilibrium configuration. Having in mind the relations \cite{wallace_47}
\begin{equation} 
\sigma_{\alpha}=\frac{1}{V_{0}}\frac{\partial E(\varepsilon)}{\partial\varepsilon_{\alpha}}\bigg|_{\varepsilon=0}
\label{eq:equ18}
\end{equation}
and
\begin{equation}
C_{\alpha \beta}=\frac{1}{V_{0}}\frac{\partial^{2}E(\varepsilon)}{\partial\varepsilon_{\alpha}\partial\varepsilon_{\beta}}\bigg|_{\varepsilon=0},
\label{eq:equ19}
\end{equation}
(\ref{eq:equ17}) equation may be written in the form of
\begin{equation} 
 E(\varepsilon)=E_{0}+V_{0}\displaystyle\sum\limits_{\alpha=1}^{6}\sigma_{\alpha}\varepsilon_{\alpha}+\frac{V_{0}}{2}\displaystyle\sum\limits_{\alpha,\beta=1}^{6}C_{\alpha\beta}\varepsilon_{\alpha}\varepsilon_{\beta}+\dots,
 \label{eq:equ20}
\end{equation}
where $V_{0}$ denotes the volume of the unstrained unit cell. If the crystalline structure is fully relaxed, the linear term in (\ref{eq:equ20}) is equal to zero and the second-order elastic constants can be obtained by evaluating total energy as a function of the applied strains. However, in order to remain in the linear regime, the deformation of the crystal should be sufficiently small. In this work, the magnitude of the applied strains $\delta$ was varied between $-0.03$ and 0.03 with a step of 0.005 (with several exceptions for tetragonal and orthorhombic phases, see Sec. \ref{sec4} B) for 9 different deformations (see Table \ref{tab1}) that correspond to the appropriate elastic constants or their combinations. The energy-strain curve of the deformed structures (\ref{eq:equ20}) was fitted with polynomials in order to extract the coefficient of a quadratic term $\delta^{2}$ which possesses the required data for the evaluation of elastic constants. For a better stability, the polynomials up to order 5 were used during the fitting procedure. While computing the total energy under deformations, the internal atomic positions were allowed to relax with the deformed cell shape and volume remaining fixed.

\begin{table}%[H] add [H] placement to break table across pages
\footnotesize
\renewcommand{\arraystretch}{0.88}
\caption{\label{tab1}
Combination of applied strains and associated space group symmetry, number of $k$ points in the irreducible Brillouin zone, and coefficient of a quadratic term in the polynomial fitting of the total energy-strain curve. Space group symmetry under deformation was identified with FINDSYM program \cite{findsym_78}.  }
 \begin{ruledtabular}
 \begin{tabular}{ccccc}
 \begin{tabular}{c}Strain\\tensor $\varepsilon$ \end{tabular}& Initial symmetry & Symmetry under deformation & \begin{tabular}{c} No. of $k$\\points \end{tabular} & \begin{tabular}{c}Coefficient \\of $\delta^2$ \end{tabular} \\
\hline
$ \left[ \begin{array}{ccc}
\delta & 0 & 0 \\
0 & 0 & 0 \\
0 & 0 & 0
\end{array} \right] $ &  \begin{tabular}{c} Cubic, No. 221 ($Pm\bar{3}m$) \\Tetr., No. 140 ($I4/mcm$) \\ Orth., No. 62 ($Pbnm$)\end{tabular}  &  \begin{tabular}{c} Tetr., No. 123 ($P4/mmm$) \\Orth., No. 72 ($Ibam$) \\ Orth., No. 62 ($Pbnm$)\end{tabular}    &  \begin{tabular}{c} 196 \\428 \\ 125  \end{tabular}&  $\frac{V_{0}}{2}C_{11}$  \\
$ \left[ \begin{array}{ccc}
0 & 0 & 0 \\
0 & \delta & 0 \\
0 & 0 & 0
\end{array} \right] $ &  Orth., No. 62 ($Pbnm$) &  Orth., No. 62 ($Pbnm$)   &  125   &  $\frac{V_{0}}{2}C_{22}$  \\
$ \left[ \begin{array}{ccc}
0 & 0 & 0 \\
0 & 0 & 0 \\
0 & 0 & \delta
\end{array} \right] $ &  \begin{tabular}{c} Tetr., No. 140 ($I4/mcm$) \\ Orth., No. 62 ($Pbnm$) \\ \end{tabular}  &  \begin{tabular}{c} Tetr., No. 140 ($I4/mcm$)  \\ Orth., No. 62 ($Pbnm$)\end{tabular}    &  \begin{tabular}{c} 244 \\ 125  \end{tabular}   &  $\frac{V_{0}}{2}C_{33}$  \\
$ \left[ \begin{array}{ccc}
\delta & 0 & 0 \\
0 & \delta& 0 \\
0 & 0 & 0
\end{array} \right] $ & \begin{tabular}{c} Cubic, No. 221 ($Pm\bar{3}m$) \\Tetr., No. 140 ($I4/mcm$) \\ Orth., No. 62 ($Pbnm$)\end{tabular}  &  \begin{tabular}{c}Tetr., No. 123 ($P4/mmm$) \\Tetr., No. 140 ($I4/mcm$) \\ Orth., No. 62 ($Pbnm$)\end{tabular}    &  \begin{tabular}{c} 301 \\244 \\ 125  \end{tabular}   &  $\frac{V_{0}}{2}(2C_{12}+C_{11}+C_{22})$  \\
$ \left[ \begin{array}{ccc}
\delta & 0 & 0 \\
0 & 0 & 0 \\
0 & 0 & \delta
\end{array} \right] $ &  \begin{tabular}{c} Tetr., No. 140 ($I4/mcm$) \\ Orth., No. 62 ($Pbnm$) \\ \end{tabular}  &  \begin{tabular}{c} Orth., No. 72 ($Ibam$)  \\ Orth., No. 62 ($Pbnm$)\end{tabular}    &  \begin{tabular}{c} 428 \\ 125  \end{tabular}   &   $\frac{V_{0}}{2}(2C_{13}+C_{11}+C_{33})$  \\
$ \left[ \begin{array}{ccc}
0 & 0 & 0 \\
0 & \delta & 0 \\
0 & 0 & \delta
\end{array} \right] $ &  Orth., No. 62 ($Pbnm$) &  Orth., No. 62 ($Pbnm$)   &  125   &  $\frac{V_{0}}{2}(2C_{23}+C_{22}+C_{33})$  \\
$ \left[ \begin{array}{ccc}
0 & 0 & 0 \\
0 & 0 & \delta \\
0 & \delta & 0
\end{array} \right] $ &  \begin{tabular}{c} Cubic, No. 221 ($Pm\bar{3}m$) \\Tetr., No. 140 ($I4/mcm$) \\ Orth., No. 62 ($Pbnm$)\end{tabular}  &  \begin{tabular}{c}Orth., No. 65 ($Cmmm$) \\Mon., No. 15 ($C/2c$) \\ Mon., No. 14 ($P2_{1}/c$)\end{tabular}    &  \begin{tabular}{c} 301 \\744 \\ 170  \end{tabular}   &  $2V_{0}C_{44}$  \\
$ \left[ \begin{array}{ccc}
0 & 0 & \delta \\
0 & 0 & 0 \\
\delta & 0 & 0
\end{array} \right] $ &  Orth., No. 62 ($Pbnm$) &  Mon., No. 14 ($P2_{1}/c$)   &  170   &  $2V_{0}C_{55}$  \\
$ \left[ \begin{array}{ccc}
0 & \delta & 0 \\
\delta & 0 & 0 \\
0 & 0 & 0
\end{array} \right] $ &   \begin{tabular}{c} Tetr., No. 140 ($I4/mcm$) \\ Orth., No. 62 ($Pbnm$) \\ \end{tabular}  &  \begin{tabular}{c} Orth., No. 69 ($Fmmm$)  \\ Mon., No. 14 ($P2_{1}/c$)\end{tabular}    &  \begin{tabular}{c} 428 \\ 170  \end{tabular}    &  $2V_{0}C_{66}$  \\
\end{tabular}
\end{ruledtabular}
\end{table}

\subsection{Mechanical stability and macroscopic elastic parameters}

A solid in equilibrium is mechanically stable if under arbitrary but small deformations the quadratic term in (\ref{eq:equ20}) is positive definite for all real values of strains unless the strains are zero. This determines restrictions, expressed in the form of inequalities, for the elastic constants \cite{grimvall_46}. For the cubic, tetragonal, and orthorhombic crystalline structures the necessary conditions for the mechanical stability are given by \cite{beckstein_2001_48}:
\begin{equation} 
C_{11}>0, \text{  } C_{44}>0, \text{  } (C_{11}-C_{12})>0, \text{  } (C_{11}+2C_{12})>0;
\label{eq:equ21}
\end{equation}
\begin{equation} 
\begin{split}
&C_{11}>0, \text{ } C_{33}>0, \text{ } C_{44}>0, \text{ } C_{66}>0, \\
(C_{11}-C_{12})>0, \text{ } &(C_{11}+C_{33}-2C_{13})>0, \text{ } (2C_{11}+C_{33}+2C_{12}+4C_{13})>0;
\end{split}
\label{eq:equ22}
\end{equation}
\begin{equation}
\begin{split}
 &C_{11}>0, \text{ } C_{22}>0, \text{ } C_{33}>0, \text{ } C_{44}>0, \text{ } C_{55}>0, \text{ } C_{66}>0, \\
(C_{11}+C_{22}-&2C_{12})>0, \text{ } (C_{11}+C_{33}-2C_{13})>0, \text{ } (C_{22}+C_{33}-2C_{23})>0, \\ &(C_{11}+C_{22}+C_{33}+2C_{12}+2C_{13}+2C_{23})>0;
\end{split}
\label{eq:equ23}
\end{equation}
respectively.

Since elastic properties of materials can be described in different ways, different relations between single-crystal and polycrystalline parameters exist. As was suggested by Voigt \cite{voigt_x}, the polycrystalline bulk ($B$) and shear ($G$) moduli can be expressed in the appropriate combinations of single-crystal elastic constants $C_{\alpha\beta}$:
\begin{equation} 
B_{V}=\frac{1}{9}[C_{11}+C_{22}+C_{33}+2(C_{12}+C_{13}+C_{23})],
\label{eq:equ24}
\end{equation}
\begin{equation} 
G_{V}=\frac{1}{15}[C_{11}+C_{22}+C_{33}-C_{12}-C_{13}-C_{23}+3(C_{44}+C_{55}+C_{66})].
\label{eq:equ25}
\end{equation}
Analogously, Reuss \cite{reuss_y} has derived the bulk and shear moduli expressions in terms of compliance constants $S_{\alpha\beta}$:
\begin{equation} 
B_{R}=[S_{11}+S_{22}+S_{33}+2(S_{12}+S_{13}+S_{23})]^{-1},
\label{eq:equ26}
\end{equation}
\begin{equation} 
G_{R}=15[4(S_{11}+S_{22}+S_{33}-S_{12}-S_{13}-S_{23})+3(S_{44}+S_{55}+S_{66})]^{-1}.
\label{eq:equ27}
\end{equation}
Voigt has based his formulation on the assumption that the strain is uniform throughout the sample, whereas the stress can be discontinuous. Reuss, on the contrary, has assumed that the stress is uniform allowing the strain to be discontinuous. Since in the first model the forces between the grains will not be in equilibrium, and in the second the distorted grains will not fit together, Hill \cite{hill_z} has shown that, for any crystalline structure, the assumptions of Voigt and Reuss lead to an upper and lower bounds of $B$ and $G$, respectively. In solid-state physics, it is common to use the arithmetic average of Voigt and Reuss bounds for the evaluation of $B$ and $G$. It is called the Voigt-Reuss-Hill (VRH) approximation:
\begin{equation} 
X_{VRH}=\frac{1}{2}(X_{V}+X_{R}), \text{ } X\equiv B,G.
\label{eq:equ28} 
\end{equation}
Regarding the general relations between Poisson's ratio ($\nu$), bulk, shear, and Young's ($Y$) moduli \cite{grimvall_46}, the VRH approximation for $\nu$ and $Y$ can be written in the form of
\begin{equation} 
\nu_{VRH}=\frac{3B_{VRH}-2G_{VRH}}{2(3B_{VRH}+G_{VRH})},
\label{eq:equ29}
\end{equation}
\begin{equation} 
Y_{VRH}=\frac{9B_{VRH}G_{VRH}}{3B_{VRH}+G_{VRH}}.
\label{eq:equ30}
\end{equation}
Alternatively, the bulk modulus can be evaluated by fitting the total energy as a function of volume to the third order Birch-Murnaghan (BM) equation of state:
\begin{equation} 
E(V)=E_{0}+\frac{9V_{0}B}{16}\left(\left[ \left(\frac{V_{0}}{V}\right)^{2/3}-1 \right]^{3}B'+ \left[ \left(\frac{V_{0}}{V}\right)^{2/3}-1 \right]^{2}\left[6-4 \left(\frac{V_{0}}{V}\right)^{2/3} \right]\right),
\label{eq:equ31}
\end{equation}
in which $E_{0}$ and $V_{0}$ denote the total energy and volume of equilibrium configuration, respectively, whereas $V$ is the volume of deformed structure and $B'$ stands for the pressure derivative of bulk modulus. The procedure is based on the constant volume but full geometry optimization under $\varepsilon_{11}=\varepsilon_{22}=\varepsilon_{33}=\delta$ (remaining $\varepsilon_{kl}=0$) deformations of the unit cell. For each phase of SrRuO$_3$, 11 points were fitted within the range of $0.92 - 1.08$ variation of the initial volume $V_0$. Since in this study all the aforementioned macroscopic elastic parameters by default are presented using VRH approximation, the bulk modulus and its pressure derivative obtained from the BM equation of state will be correspondingly denoted as $B_{BM}$ and $B'_{BM}$.

For polycrystalline material, the longitudinal ($v_{L}$) and transverse ($v_{T}$) sound velocities can be written in terms of bulk and shear moduli obtained within the VRH approximation \cite{grimvall_46}:
\begin{equation} 
v_{L}=\sqrt{\frac{3B_{VRH}+4G_{VRH}}{3\varrho}},
\label{eq:equ32}
\end{equation}
\begin{equation} 
v_{T}=\sqrt{\frac{G_{VRH}}{\varrho}},
\label{eq:equ33}
\end{equation}
where $\varrho$ denotes the mass density. The Debye temperature then has the form
\begin{equation} 
\theta_{D}=\frac{\hbar \bar{v}}{k_{B}}\left(\frac{6\pi^{2}N}{V_{0}} \right)^{1/3},
\label{eq:equ34}
\end{equation}
where
\begin{equation} 
\bar{v}=\left(\frac{1}{3}\left(\frac{1}{v^{3}_{L}}+\frac{2}{v^{3}_{T}} \right) \right)^{-1/3}
\label{eq:equ35}
\end{equation}
stands for an average sound velocity, $N$ is the number of atoms per unit cell, and $V_0$ is its volume.

\section{Computational details}

The first-principles periodic calculations were carried out using CRYSTAL09 (Ref. \onlinecite{crystal09_49}) code, in which the crystalline wave functions are expanded as a linear combination of atom-centered Gaussian orbitals (LCAO). Equilibrium atomic positions and lattice constants were obtained using analytical gradients of the total energy with respect to atomic coordinates and unit-cell parameters within the quasi-Newton algorithm. The Hessian matrix containing numerical second derivatives of the total energy was updated by means of the Broyden-Fletcher-Goldfarb-Shanno (BFGS) scheme. The initial values of Hessian matrix were generated using a classical model proposed by Schlegel \cite{schlegel_1_50, schlegel_2_51}. Optimization convergence was checked on the root-mean-square (RMS) and the absolute magnitude of the largest components of both the gradients and the estimated nuclear displacements. The optimization was considered complete when these four parameters satisfied the determined thresholds which for the RMS values of gradients and displacements were set to 0.00006 and 0.00012 a.u., respectively. The maximum allowed gradients and displacements were 1.5 times larger. In order to improve the self-consistence field convergence, the Kohn-Sham matrix mixing technique together with modified Broyden's scheme \cite{broyden_52} or Anderson's method \cite{anderson_53} were applied. The default values of truncation criteria for bielectronic integrals were modified to tighter ones by setting the overlap thresholds for exchange and Coulomb integrals to $10^{-7}$ for orthorhombic and to $10^{-9}$ for tetragonal and cubic phases. Analogously, the tolerance on change in total energy was tightened to $10^{-9}$ a.u. for tetragonal and cubic phases, whereas for the computationally heavier orthorhombic phase the parameter was set to $10^{-7}$ a.u. The number of asymmetric $k$ points in the irreducible Brillouin zone used for each of the phase is given in Table \ref{tab1}. 

In CRYSTAL09, while numerically evaluating DFT exchange-correlation contribution, grid points are generated using Gauss-Legendre radial quadrature and Lebedev two-dimensional angular point distribution. In this work, an extra large grid was used, which contains 75 radial points and 974 angular points in the most accurate integration region. Concerning the basis sets, small-core Hay-Wadt pseudopotentials \cite{hay_1985_54} (PPs) corresponding to 28 core electrons were applied for both Sr and Ru atoms. The valence part of the basis set for Sr was taken from the strontium titanate study \cite{piskunov_2004_55}. For Ru, the valence functions combined with the Hay-Wadt PP were adopted from the modified LANL2DZ basis set \cite{lanl2dz_56} available online at EMSL database \cite{lanl2dz_http_57}. In order to avoid numerical problems usually caused by too diffuse functions (since they tend to overlap with the core functions of the neighboring atoms), the exponents smaller than 0.08 bohr$^{-2}$ were removed. The outermost $p$ shell exponent $\alpha_{p} = 0.083$ bohr$^{-2}$ was decontracted and, together with the most external $d$ shell exponent $\alpha_{d} = 0.1501$ bohr$^{-2}$, optimized by attempting to minimize the total energy per unit cell with the lattice constants and atomic positions fixed at experimental values. After optimization, the corresponding $p$ and $d$ exponents were set to 0.1301 and 0.1329 bohr$^{-2}$, respectively. For O, the all-electron basis set with a double set of $d$ functions was taken from Ref. \onlinecite{valenzano_2006_58}. Basis sets for Sr and O atoms are also available online at CRYSTAL Basis Sets Library \cite{crysta09_49a}.  

\section{Results and discussion}
\label{sec4}

\subsection{Lattice constants}

Since the accuracy of the predicted lattice constants is a test of the accuracy of the density functional for the description of the exchange-correlation energy \cite{hao_2012_59}, in the first place we present the equilibrium lattice constants calculated using different DFT approaches. They are shown in Table \ref{tab2}, \ref{tab3}, and \ref{tab4} for cubic, tetragonal, and orthorhombic phases of SrRuO$_{3}$, respectively. The available experimental data is also given therein, however no thermal and/or zero-point anharmonic expansion (ZPAE) corrections necessary for the precise comparison between theory and experiment were applied. On one hand, the extrapolation from 973 (cubic phase) or 823 (tetragonal phase) to 0 K using an average linear thermal expansion coefficient would not be accurate and in principle should substantially exceed the effect of ZPAE. Thus, the evaluation of the ZPAE correction alone would neither improve the situation for any of these two phases. On the other hand, the ZPAE correction, most commonly written as Eq. (A6) in Ref. \onlinecite{alchagirov_2001_60}, is applicable only for the cubic systems meaning that the low-temperature experimental data for orthorhombic SrRuO$_{3}$ can not be simply corrected. Although it is clear that the presented calculations and experimental values should not be compared in a very strict manner, some important trends can still be observed. Having in mind that the combination of thermal effects and ZPAE usually expand lattice constant compared to its value at 0 K, it is evident that PBE tends to overestimate the lattice parameters for all three phases of SrRuO$_{3}$. The opposite trend is seen for LDA, most obviously for the orthorhombic SrRuO$_{3}$. In order for the LDA results to perfectly match the experimental data, the ZPAE correction should reach $\sim$0.75\% of a lattice constant. This scenario is not realistic, since ZPAE can expand the equilibrium lattice constant by 1\% for light atoms like Li and much less for heavy atoms \cite{hao_2012_59}. Moreover, the evaluation of ZPAE correction for cubic phase of SrRuO$_{3}$ reveals that with parameters obtained from different DFT approximations it does not exceed $\sim$0.13\%. Therefore, it is very likely that LDA and PBE approaches tend to correspondingly underestimate and overestimate lattice constants for all three phases of SrRuO$_{3}$. This observation is strengthened by the fact that calculated values of the revised functionals fall within the range between LDA and PBE. It is a well known tendency noticed for a large set of solids \cite{haas_2009_24}. Among the revised functionals, SOGGA gives the lowest, whereas WC gives the highest values for all three phases. However, the difference between the lattice constants is quite small and does not exceed $\sim$0.34\%, compared to the largest difference of $\sim$1.8\% between SOGGA and PBE and the one of $\sim$1.05\% between WC and LDA. The values of PBEsol and PBEsol\textsuperscript{PBE} fall within the inner range between SOGGA and WC. According to the similar performance of the revised functionals, they can be divided into two groups, with SOGGA and PBEsol\textsuperscript{PBE} being in one and PBEsol and WC being in another. On the whole, if one would make an assumption that the ZPAE correction value of $\sim$0.13\% could be directly applied for the orthorhombic phase of SrRuO$_{3}$, then all three lattice parameters $a$, $b$, and $c$ calculated using all the revised density functionals under study here would have a deviation smaller than 0.5\% from the ZPAE-corrected experimental data. That would correspond to the ``good" theoretical values. Meanwhile, neither LDA, nor PBE could satisfy this condition despite LDA being noticeably closer to the experiment than PBE. Interestingly, without the application of ZPAE correction the revised functionals would satisfy even a tighter deviation of 0.3\%. These observations apparently indicate the effectiveness of the revision made to the construction of the employed density functionals. 

 \begin{table}%[H] add [H] placement to break table across pages
 \caption{\label{tab2}
Equilibrium lattice constant $a$ (in \AA), elastic constants $C_{\alpha\beta}$ (in GPa), bulk modulus $B$ (in GPa) from VRH approximation, bulk modulus $B_{BM}$ (in GPa) and its pressure derivative $B'_{BM}$ from BM equation of state, shear modulus $G$ (in GPa), Young's modulus $Y$ (in GPa), Poisson's ratio $\nu$, longitudinal sound velocity $v_L$ (in m/s), transverse sound velocity $v_T$ (in m/s), average sound velocity $\bar{v}$ (in m/s), and Debye temperature $\theta_D$ (in K) calculated and compared to the available experimental data for cubic SrRuO$_3$.      }
 \begin{ruledtabular}
 \begin{tabular}{cccccccc}
    & LDA & SOGGA & PBEsol\textsuperscript{PBE} & PBEsol & WC & PBE & Expt. \\
\hline
 $a$   & 3.900 & 3.924 & 3.926 & 3.933 & 3.936 & 3.982 & 3.955\footnote{Ref. \onlinecite{kennedy_1998_12} at 973 K.}, 3.974\footnote{Ref. \onlinecite{chakoumakos_1998_13} at 1082 K.} \\
 $C_{11}$   & 240.2 & 210.3 & 208.7 & 205.4 & 205.6 & 176.9 &  \\
 $C_{12}$   & 204.6 & 194.6 & 193.2 & 192.9 & 190.2 & 176.6 &  \\
 $C_{44}$   & 76.1 & 73.8 & 73.7 & 73.4 & 73.1 & 70.9 &  \\
 $B$   & 216.5 & 199.8 & 198.4 & 197.1 & 195.3 & 176.8 &  \\
 $B_{BM}$   & 215.9 & 198.9 & 197.6 & 196.3 & 196.2 & 176.5 &  \\
 $B'_{BM}$   & 4.55 & 4.26 & 4.30 & 4.34 & 4.58 & 4.41 &  \\
 $G$   & 42.9 & 32.1 & 32.0 & 30.2 & 31.8 & 21.1 &  \\
 $Y$   & 120.6 & 91.6 & 91.2 & 86.2 & 90.4 & 60.7 &  \\
 $\nu$   & 0.407 & 0.424 & 0.423 & 0.427 & 0.423 & 0.443 &  \\
 $v_L$   & 6415 & 6095 & 6082 & 6047 & 6058 & 5724 &  \\
 $v_T$   & 2539 & 2219 & 2217 & 2157 & 2215 & 1835 &  \\
 $\bar{v}$   & 2877 & 2520 & 2518 & 2451 & 2515 & 2089 &  \\
 $\theta_D$   & 375.5 & 326.9 & 326.3 & 317.2 & 325.4 & 267.1 &  \\
 \end{tabular}
 \end{ruledtabular}
 \end{table}

As was mentioned in the Introduction, several theoretical works that include calculated structural parameters of cubic and orthorhombic SrRuO$_{3}$ can be found in the literature. Naturally, due to the closest resemblance to the low-temperature experimental data the most intriguing is the one that incorporates the revised functionals. The fact that the same computer code was employed in the present paper and Ref. \onlinecite{garcia_2012_36} makes the comparison rather straightforward. Nevertheless, it is surprising to note that all the lattice constants calculated with PBEsol and WC functionals are somewhat larger in Ref. \onlinecite{garcia_2012_36} than in this work despite the tighter convergence criteria applied therein. By taking into account that experimental parameters can only be smaller after the ZPAE correction, one can conclude that our calculations exhibit a better agreement with the low-temperature experiment. The largest mismatch between the same revised functionals reaches up to $\sim$0.7\% meaning that not all the lattice parameters estimated in Ref. \onlinecite{garcia_2012_36} correspond to the ``good" theoretical values. Moreover, the ZPAE correction could make the situation even worse. The possible source of the following discrepancy probably lies in the basis sets utilized in the calculations. Since basis set for Sr is the same in both studies and basis set for O is of the same quality, the factor that remains is the difference in basis sets for Ru. From this observation it is obvious that basis set for a particular element optimized in a system under consideration can become an appreciable advantage. It is interesting to notice that the inclusion of the HF exchange energy partly compensates for the observed mismatch. Although the incorporation of the hybrid scheme also results in overall a better description of tilting and rotation angles, the lattice parameters can be easily overcorrected by increasing the amount of HF exchange. Having in mind that calculations are basis set dependent, the exact amount of hybrid mixing that should be applied becomes unclear. Because of clear-cut physics that stands beyond the revised functionals, we prefer them over the hybrids for the description of SrRuO$_{3}$. 

 \begin{table}%[H] add [H] placement to break table across pages 
 \caption{\label{tab3}
Equilibrium lattice constant $a$ and $c$ (in \AA), elastic constants $C_{\alpha\beta}$ (in GPa), bulk modulus $B$ (in GPa) from VRH approximation, bulk modulus $B_{BM}$ (in GPa) and its pressure derivative $B'_{BM}$ from BM equation of state, shear modulus $G$ (in GPa), Young's modulus $Y$ (in GPa), Poisson's ratio $\nu$, longitudinal sound velocity $v_L$ (in m/s), transverse sound velocity $v_T$ (in m/s), average sound velocity $\bar{v}$ (in m/s), and Debye temperature $\theta_D$ (in K) calculated and compared to the available experimental data for tetragonal SrRuO$_3$.      }
 \begin{ruledtabular}
 \begin{tabular}{cccccccc}
    & LDA & SOGGA & PBEsol\textsuperscript{PBE} & PBEsol & WC & PBE & Expt. \\
\hline
 $a$   & 5.513 & 5.554 & 5.557 & 5.570 & 5.572 & 5.656 & 5.578\footnote{Ref. \onlinecite{kennedy_1998_12} at 823 K.}, 5.589\footnote{Ref. \onlinecite{chakoumakos_1998_13} at 894 K.} \\
 $c$   & 7.773 & 7.801 & 7.805 & 7.813 & 7.820 & 7.872 & 7.908\footnotemark[1], 7.917\footnotemark[2] \\
 $C_{11}$   & 268.9 & 250.5 & 249.1 & 246.7 & 245.8 & 223.8 &  \\
 $C_{12}$   & 146.7 & 132.9 & 131.8 & 129.9 & 129.7 & 110.6 &  \\
 $C_{13}$   & 185.9 & 172.4 & 171.9 & 170.7 & 170.6 & 151.9 &  \\
 $C_{33}$   & 343.8 & 325.1 & 322.6 & 321.7 & 321.1 & 299.9 &  \\
 $C_{44}$\footnote{calculated within a range of strains $\delta$ from -0.0003 to 0.0003.}  & 71.6 & 68.8 & 68.7 & 68.8 & 68.6 & 65.7 &  \\
 $C_{66}$   & 99.4 & 91.1 & 90.3 & 88.1 & 87.7 & 75.6 &  \\
 $B$   & 209.2 & 193.7 & 192.7 & 190.9 & 190.5 & 170.5 &  \\
 $B_{BM}$   & 205.1 & 189.4 & 187.9 & 186.2 & 185.8 & 165.3 &  \\
 $B'_{BM}$   & 4.83 & 4.89 & 4.81 & 4.82 & 4.85 & 5.19 &  \\
 $G$   & 71.2 & 67.7 & 67.3 & 66.8 & 66.5 & 62.8 &  \\
 $Y$   & 191.9 & 181.9 & 180.8 & 179.6 & 178.8 & 167.7 &  \\
 $\nu$   & 0.347 & 0.344 & 0.344 & 0.343 & 0.344 & 0.336 &  \\
 $v_L$   & 6746 & 6579 & 6566 & 6556 & 6553 & 6367 &  \\
 $v_T$   & 3265 & 3212 & 3205 & 3203 & 3199 & 3164 &  \\
 $\bar{v}$   & 3669 & 3608 & 3600 & 3598 & 3594 & 3551 &  \\
 $\theta_D$   & 479.6 & 468.8 & 467.4 & 466.4 & 465.5 & 454.3 &  \\
 \end{tabular}
 \end{ruledtabular}
 \end{table}

\begin{figure}
\includegraphics[scale=0.7]{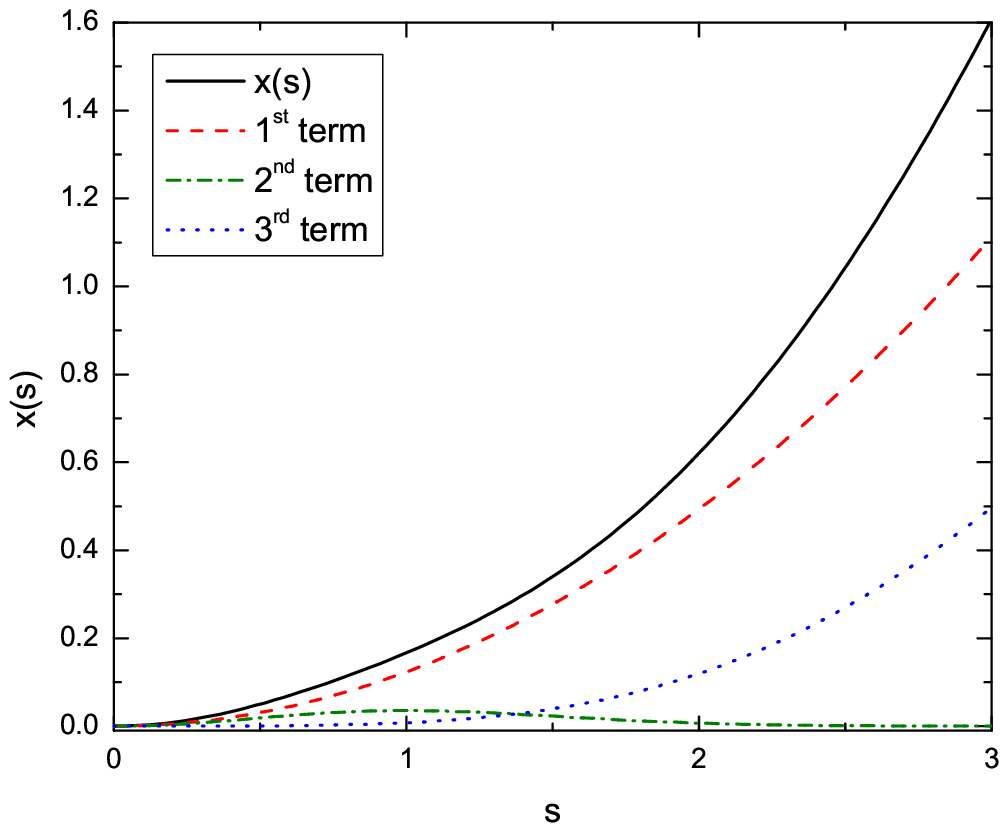}
\caption{\label{fig1}
The behavior of the terms of $x(s)$ [Eq. (\ref{eq:equ10})] introduced in the WC functional.}
\end{figure}

The fact that we used the same correlation functional for SOGGA, PBEsol\textsuperscript{PBE}, WC, and PBE allows us to present a few significant insights into the inner structure of SrRuO$_{3}$. Firstly, the step that leads from $F^{PBE}_{X}(s)$ to $F^{PBEsol}_{X}(s)$ seems to be the most important for the accurate description of the lattice parameters of SrRuO$_{3}$. This can be easily noticed by comparing the results of PBEsol\textsuperscript{PBE} and PBE, since the only difference between these two functionals is the value of parameter $\mu$, which was set from $\mu^{PBE}=0.2195$ to $\mu^{PBEsol}=10/81\approx 0.1234$. Secondly, the modification of parameter $\beta$ ($\beta^{PBE}=0.0667\rightarrow \beta^{PBEsol}=0.046$), which appears in the correlation functional, has much less impact (seen from comparison between PBEsol\textsuperscript{PBE} and PBEsol), indicating that the magnitude of the exchange energy is substantially larger compared to the correlation energy. What is more, a close look at Fig. 1 in Ref. \onlinecite{wc_20} reveals that $F^{PBE}_{X}(s)$ and $F^{WC}_{X}(s)$ are nearly identical for $s\leq 0.5$. Having in mind that the lattice constants obtained with PBE and WC functionals are noticeably different, one can make an assumption that the average reduced density gradient $s$ should exceed this value in SrRuO$_{3}$. Although the WC functional has rather complicated form, which ensures the same behavior as PBE for $s \rightarrow 0$ and $s \rightarrow \infty$, examination of the terms in $x(s)$, presented in Fig. \ref{fig1}, shows that at $s$ values larger than $\sim$0.5 the first term $\frac{10}{81}s^2$ starts dominating over the remaining two terms. Since this term is exactly the same as it is in PBEsol, it becomes clear why these two functionals having distinct $\mu$ values are both able to yield ``good" theoretical results. This once again confirms the significance of the exact second-order GE for the description of the lattice parameters of SrRuO$_{3}$.

\begin{table}
\renewcommand{\arraystretch}{0.84}
\caption{\label{tab4}
Equilibrium lattice constant $a$, $b$, and $c$ (in \AA), elastic constants $C_{\alpha\beta}$ (in GPa), bulk modulus $B$ (in GPa) from VRH approximation, bulk modulus $B_{BM}$ (in GPa) and its pressure derivative $B'_{BM}$ from BM equation of state, shear modulus $G$ (in GPa), Young's modulus $Y$ (in GPa), Poisson's ratio $\nu$, longitudinal sound velocity $v_L$ (in m/s), transverse sound velocity $v_T$ (in m/s), average sound velocity $\bar{v}$ (in m/s), and Debye temperature $\theta_D$ (in K) calculated and compared to the available experimental data for orthorhombic SrRuO$_3$.      }
 \begin{ruledtabular}
 \begin{tabular}{cccccccc}
    & LDA & SOGGA & PBEsol\textsuperscript{PBE} & PBEsol & WC & PBE & Expt. \\
\hline
 $a$   & 5.527 & 5.556 & 5.558 & 5.568 & 5.570 & 5.631 & 5.566\footnote{Ref. \onlinecite{bushmeleva_2006_16} at 1.5 K.}, 5.566\footnote{Ref. \onlinecite{kiyama_1996_17} at 12 K. } \\
 $b$   & 5.491 & 5.529 & 5.533 & 5.543 & 5.548 & 5.621 & 5.532\footnotemark[1], 5.529\footnotemark[2] \\
 $c$   & 7.779 & 7.824 & 7.828 & 7.841 & 7.846 & 7.937 & 7.845\footnotemark[1], 7.844\footnotemark[2] \\
 $C_{11}$   & 306.9 & 286.9 & 285.2 & 284.8 & 284.2 & 260.1 &  \\
 $C_{12}$   & 186.7 & 171.1 & 169.5 & 167.9 & 167.8 & 143.4 &  \\
 $C_{13}$   & 156.2 & 145.1 & 144.4 & 143.6 & 143.4 & 128.2 &  \\
 $C_{22}$   & 274.5 & 256.9 & 255.6 & 253.9 & 252.7 & 230.1 &  \\
 $C_{23}$   & 147.7 & 139.8 & 139.1 & 138.3 & 137.9 & 125.9 &  \\
 $C_{33}$   & 338.6 & 312.6 & 310.8 & 307.3 & 306.8 & 272.1 &  \\
 $C_{44}$\footnote{calculated within a range of strains $\delta$ from -0.015 to 0.015.}  & 72.7 & 71.4 & 71.9 & 72.2 & 72.4 & 71.1 &  \\
 $C_{55}$   & 67.7 & 66.4 & 66.4 & 66.7 & 66.7 & 65.9 &  \\
 $C_{66}$   & 88.1 & 80.9 & 80.1 & 78.7 & 78.4 & 67.3 &  \\
 $B$   & 210.8 & 196.1 & 194.9 & 193.6 & 193.1 & 172.7 &  \\
 $B_{BM}$   & 210.5 & 195.9 & 194.4 & 192.8 & 192.5 & 172.8 & 192.3\footnote{Ref. \onlinecite{hamlin_2007_38} at room temperature.} \\
 $B'_{BM}$   & 4.75 & 4.64 & 4.67 & 4.65 & 4.66 & 4.69 & 5.03\footnotemark[4] \\
 $G$   & 72.8 & 69.3 & 69.1 & 68.9 & 68.7 & 64.6 & 60.1\footnote{Ref. \onlinecite{yamanaka_2004_14} at room temperature.} \\
 $Y$   & 195.8 & 185.9 & 185.4 & 184.7 & 184.3 & 172.3 & 161\footnotemark[5] \\
 $\nu$   & 0.345 & 0.342 & 0.341 & 0.341 & 0.341 & 0.334 &  \\
 $v_L$   & 6782 & 6625 & 6614 & 6613 & 6612 & 6416 & 6312\footnotemark[5] \\
 $v_T$   & 3298 & 3246 & 3246 & 3249 & 3248 & 3206 & 3083\footnotemark[5] \\
 $\bar{v}$   & 3706 & 3646 & 3645 & 3648 & 3648 & 3596 &  \\
 $\theta_D$   & 484.5 & 472.8 & 473.5 & 473.1 & 472.7 & 460.6 & 448\footnotemark[5], 457\footnote{Ref. \onlinecite{dabrowski_2005_37} from fitting in the temperature range of 163-300 K.} \\ 
\end{tabular}
\end{ruledtabular}
\end{table}

Another close look at Fig. 1 in Ref. \onlinecite{sogga_19} indicates that the SOGGA and PBEsol $F_{X}(s)$ curves are almost identical up to $s\leq1.5$. This observation is perfectly consistent with the results of SOGGA and PBEsol\textsuperscript{PBE} which due to very close performance were assigned to the same group. If the average value of $s$ would be higher, we should obtain a more pronounced difference between SOGGA and PBEsol\textsuperscript{PBE}. However, the difference is negligible, meaning that the average value of $s$ should fall in the range of $0.5\leq s \leq 1.5$. Such range testifies about a moderately-varying density in SrRuO$_{3}$. It is interesting to note that another group of functionals - PBEsol and WC - also yielding very similar results have different parameters in exchange and correlation parts. As we have already found out that the first term in $x(s)$ is mainly responsible for the performance of WC, one can note that the effect of the addition of remaining two terms somewhat corresponds to the effect of the modification of correlation functional in PBEsol. It is also worth mentioning that the $F^{SOGGA}_{X}(s)$, which is taken as a half-and-half mixing of the PBE and RPBE exchange functionals, would produce larger lattice constants if the parameter $\kappa$ remains the same as in PBE ($\kappa^{PBE}=0.804$). The tighter value of 0.552 somehow compensates for this increase resulting in almost identical behavior to PBEsol up to $s\leq 1.5$. 

From all above we can conclude that regarding the accuracy of the lattice constants no clear winner has emerged among the revised functionals. It is very likely that all the revised functionals applied in this work should give ``good" theoretical values, at least for the orthorhombic phase of SrRuO$_{3}$. A small deviation between the calculated results shows that the change in the form of the exchange enhancement factor (WC and SOGGA) or variation of parameter $\beta$ (PBEsol) does not have a noticeable influence. Therefore, the PBE-like functionals with the parameter $\mu$ restored back to 10/81 can be considered as simple but reliable tool for the investigation of the crystalline structure of SrRuO$_{3}$. For this reason, we choose PBEsol as a default functional for the demonstration of SrRuO$_{3}$ behavior and properties throughout this study.       

\subsection{Elastic properties and mechanical stability}

Due to the lack of experimental as well as theoretical data, we can only make a comparison between our own elastic constants obtained within different DFT approaches. The tendency that can be observed is exactly opposite to the one found for the lattice constants. In this case, LDA gives the highest, whereas PBE gives the lowest values of $C_{\alpha\beta}$ for all three phases of SrRuO$_{3}$. The revised functionals perform quite similarly giving the values that fall between the range determined by LDA and PBE. This result once more confirms that such quantities as elastic constants and polycrystalline moduli crucially depend on reliable evaluation of the lattice parameters.  

\begin{figure}
\includegraphics[scale=0.7]{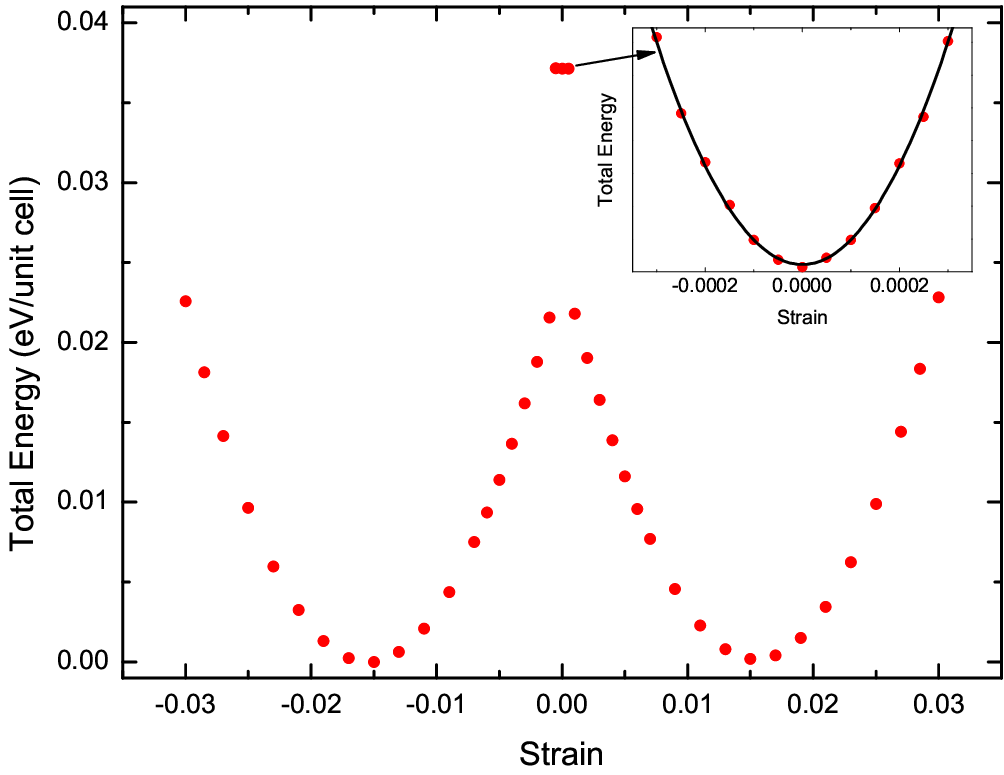}
\caption{\label{fig2}
Total energy of ground-state tetragonal SrRuO$_{3}$ as a function of strain under $C_{44}$ related shear deformation. The inset demonstrates the usual parabolic behavior when the magnitude of applied strains is reduced from -0.0003 to 0.0003. }
\end{figure}

Concerning the values of $C_{44}$ for tetragonal and orthorhombic phases of SrRuO$_{3}$, a few issues must be clarified. Firstly, our calculations for tetragonal structure show that strains $\delta$ taken in steps of 0.005 from -0.03 to 0.03 induce internal forces that drive lattice to the energetically lower states compared to the unstrained crystal (see Fig. \ref{fig2}). Neither different DFT approaches, nor modification of calculation parameters have appreciable influence on this result. This basically means that tetragonal SrRuO$_{3}$ is mechanically unstable at zero temperature (and zero pressure \cite{comment_61}) in spite of the fact that remaining $C_{\alpha\beta}$ satisfy all the stability conditions given by Eq. (\ref{eq:equ22}). In other words, under ordinary $C_{44}$ related deformation a more stable monoclinic structure emerges resulting in a negative value of $C_{44}$ which in turn characterizes a pure shear instability. The following finding is in agreement with the fact that bulk samples of tetragonal phase have not been experimentally observed at low temperatures. Yet, it is worth to note that room-temperature stabilized tetragonal SrRuO$_{3}$ has been successfully obtained by tensile biaxial strain when grown as thin film on various substrates \cite{vailionis_2008_62, chang_2011_63}. We would like to emphasize that these results are not in contradiction with our data, since the stability of thin film is determined by different factors compared to the bulk specimens \cite{wang_2003_64}. However, a question of what exactly makes the tetragonal phase experimentally observable at high temperatures requires a separate study, therefore this interesting issue lies outside the scope of the present paper. 

By remembering that the elastic energy, written as quadratic term in (\ref{eq:equ20}), must be positive definite under any arbitrary but small deformations, we were curious to find out whether the limit at which the system can be described by the usual parabolic behavior exists. For this purpose, we had to narrow the range of the applied strains and to tighten the convergence criteria. The first attempt to reduce the magnitude of the applied strains from -0.003 to 0.003 was not successful, since the total energy of the strained lattice was still lower than that of the unstrained one. However, the second attempt in which the magnitude of the applied strains was lowered from -0.0003 to 0.0003 appeared to be fruitful (see inset in Fig. \ref{fig2}). Certainly, in the first place we wanted to make sure that such calculations with very small changes in the total energy were not overpowered by numerical fluctuations and thus could be treated seriously. For the comparison, we tried to evaluate $C_{44}$ without the geometry optimization (i. e. elastic constant of the unrelaxed structure) by varying $\delta$ from -0.03 to 0.03 and from -0.0003 to 0.0003. The obtained results for PBEsol revealed that in both cases the values of $C_{44}$ were very close to each other: 85.6 against 86.3 GPa, respectively. Despite this correspondence, it is obvious that in reality even at temperatures very close to 0 K zero-point fluctuations have a larger impact on atomic positions than such tiny strains. But since in DFT framework nuclei of the atoms form a perfectly static lattice, even the smallest atomic displacements in the system can be treated as reflecting its unique response to deformation. Thus, a purely theoretical modeling allows to analyze material's behavior that could be never observed experimentally due to manifestation of quantum effects. Having in mind the thorough analysis \cite{perger_2009_65} in which the stability of elastic constants is attained using similarly small deformations, the values of $C_{44}$ presented in Table \ref{tab3} should provide realistic information on properties of tetragonal SrRuO$_{3}$. Surely, it would be particularly interesting to find out whether our results show a correspondence with the experiment or at least finite-temperature calculations. 

\begin{figure}
\includegraphics[scale=0.7]{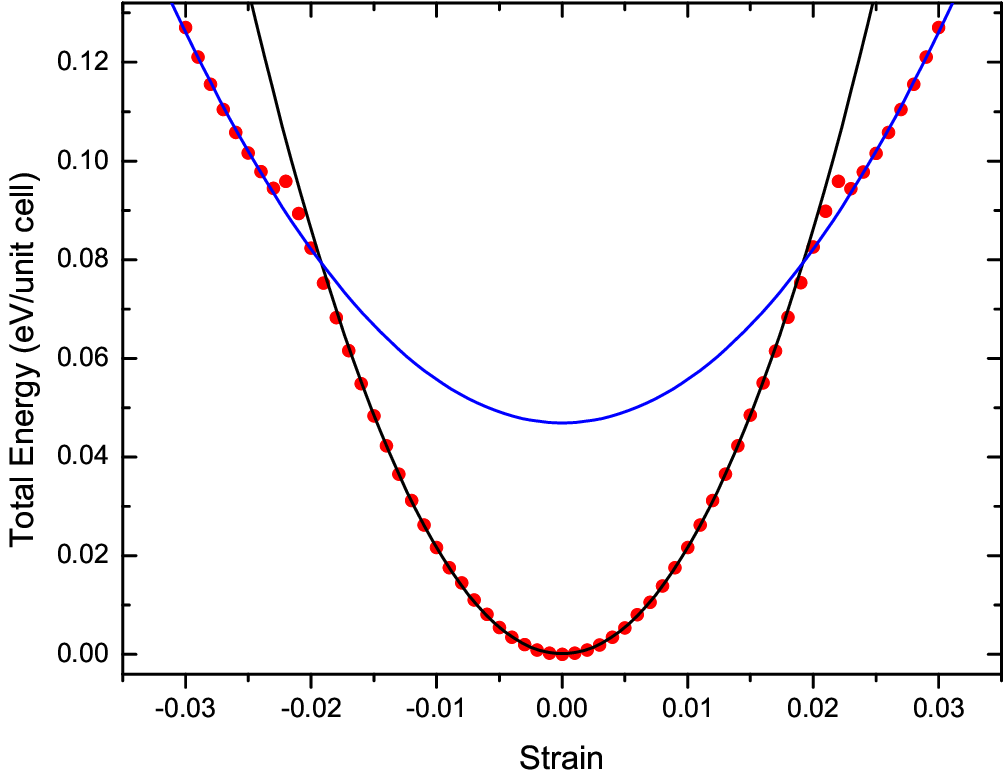}
\caption{\label{fig3}
Total energy of ground-state orthorhombic SrRuO$_{3}$ as a function of strain under $C_{44}$ related shear deformation. The lower parabola was obtained by polynomial fitting with magnitude of applied strains confined from -0.015 to 0.015. The upper parabola was fitted with strains falling in the range of $-0.03\ldots -0.023$ and $0.023\ldots 0.03$.  }
\end{figure}

Another issue is related to the calculation of $C_{44}$ for the orthorhombic phase of SrRuO$_{3}$. The application of strains taken from the standard range (from -0.03 to 0.03) has revealed a distinct deviation from the initial parabolic behavior apparently seen at strain values of $\pm$0.021 - $\pm$0.024 (depending on the DFT approach used - in case of PBEsol, see Fig. \ref{fig3}). Interestingly, the subsequent behavior of the deformed material can be approximated by another parabola with quadratic coefficient being roughly two times smaller. It indicates that the strain-induced structural transformation causes quite substantial softening of $C_{44}$. The fact that the crystallographic space group of the system remained unchanged throughout the whole range of applied strains makes us firmly believe that we are dealing with isosymmetric phase transition, which is a fairly rare phenomenon for crystalline materials. In order to get a more accurate picture of isosymmetric SrRuO$_{3}$ behavior, we have increased the number of strains from standard 12 to 60. A close examination of calculations showed that a small deviation from the initial parabola begins to emerge at $\pm$0.017. Therefore, for the precise evaluation of $C_{44}$, we have confined the range of strains from -0.015 to 0.015. Thus obtained values of $C_{44}$ are presented in Table \ref{tab4}.

Despite the unusual behavior of orthorhombic SrRuO$_{3}$, the mechanical stability of this phase is beyond doubt, since all the elastic constants   calculated within the standard range of strains obey the stability criteria expressed by Eq. (\ref{eq:equ23}). The same statement holds for the cubic phase of SrRuO$_{3}$ because its mechanical stability conditions [Eq. (\ref{eq:equ21})] are also satisfied. However, the stability restrictions do not tell much about the relative magnitudes of elastic constants which in general reflect the strength of the interatomic forces in the solid. For example, the comparison between tetragonal and orthorhombic phases indicates that $C_{11}^{Tetr.}=C_{22}^{Tetr.}<C_{33}^{Tetr.}$ and $C_{22}^{Orth.}<C_{11}^{Orth.}<C_{33}^{Orth.}$. As $C_{11}$, $C_{22}$, and $C_{33}$ are directly associated with the changes in the lattice along the $x$, $y$, and $z$ axis, respectively, one can conclude that atomic bonding along the $z$ axis is stronger than that along the $x$ or $y$ axis for both phases of SrRuO$_3$. Smaller values of $C_{44}$ and $C_{55}$ compared to $C_{66}$ mean that lattice is more easily deformed by a pure shear around the $x$ or $y$ axis in comparison to the $z$ axis. Since bulk modulus $B$ is essentially a special case of elastic constant and can be treated as a measure of the average bond strength, it is interesting to note that all three phases exhibit a very close value of it. This means that polycrystalline SrRuO$_3$ retains its level of compressibility throughout a variety of geometric configurations. The fact that the largest discrepancy between the bulk moduli obtained using VRH approximation and BM equation of state does not exceed $\sim$3\% demonstrates the reliability of performed calculations. Unlike the bulk modulus, shear and Young's moduli which describe the resistance of a material against the shear and uniaxial deformations, respectively, have significantly lower values for cubic SrRuO$_3$ compared to the tetragonal and orthorhombic phases. Due to the interdependence between the polycrystalline parameters, the opposite tendency holds for Poisson's ratio which in turn defines the lateral structural deformation when material is stretched or compressed. For high-temperature cubic SrRuO$_3$, the average value of $\nu$ is $\sim$0.42, whereas low-temperature orthorhombic SrRuO$_3$ exhibits $\sim$0.34 which is a common value for most of the metals. Lower shear modulus for cubic SrRuO$_3$ also points to lower hardness and hence the decreased resistance against plastic deformation. Despite that, a high $B/G$ ratio is inherent for all three phases of SrRuO$_3$ revealing that this material behaves in a ductile manner. The most common critical value that separates ductile and brittle nature was established by Pugh \cite{pugh_1954_66}: if $B/G>1.75$, the material behaves in a ductile manner, otherwise, its behavior should be associated with brittleness. Intriguingly, SrRuO$_3$ satisfies even tighter criterion of $B/G\sim 2.67$ proposed by Frantsevich $et$ $al$. \cite{frantsevich_67} which in a similar fashion distinguishes ductility from brittleness. It is worth mentioning that the following result is valid for all density functionals applied in this study. The fact that cubic SrRuO$_3$ becomes softer against shear-type distortions is also reflected in the lower sound velocities and Debye temperature. In the meantime, tetragonal phase is able to retain values of polycrystalline parameters that are very close to the orthorhombic ones. 

Another property that should be taken into account for the comprehensive description of the polycrystalline material is its elastic anisotropy. Having in mind that there is no unique method to determine the degree of this characteristic, in the first place we employ a concept introduced by Chung and Buessem \cite{chung_1967_68} where the percentage anisotropy in compressibility and shear is defined as $A_{B}=(B_{V}-B_{R})/(B_{V}+B_{R})$ and $A_{G}=(G_{V}-G_{R})/(G_{V}+G_{R})$, respectively. A value of zero corresponds to elastic isotropy and a value of 100\% identifies the largest possible anisotropy. Within PBEsol approximation, orthorhombic phase possesses $A_{B}\sim 0.2$\% and $A_{G}\sim 1.5$\%, while tetragonal SrRuO$_3$ acquires $A_{B}\sim 2.3$\% and $A_{G}\sim 1.8$\%. This result reveals that orthorhombic and tetragonal SrRuO$_3$ are similarly anisotropic in shear, whereas anisotropy in compressibility is more pronounced for the latter symmetry. On the whole, the values of anisotropic factors show that orthorhombic and tetragonal phases can be considered as weakly elastically anisotropic. However, completely different situation emerges for cubic SrRuO$_3$. Although in this case isotropy in compressibility is assured by $B_{V}=B_{R}$, a very large value of $A_{G}\sim 54$\% indicates a high degree of anisotropy in shear. It in turn means that cubic phase can be considered as highly elastically anisotropic. The following conclusions are also confirmed by estimation of universal elastic anisotropy index $A^{U}=5(G_{V}/G_{R})+(B_{V}/B_{R})-6$ proposed by Ranganathan and Ostoja-Starzewski \cite{universal_69}. In this analysis method, the departure of $A^{U}$ from zero defines the extent of single crystal anisotropy accounting for both the bulk and the shear contributions. For orthorhombic, tetragonal, and cubic SrRuO$_3$, the values of $A^{U}$ are $\sim$0.16, 0.23, and 11.8, respectively, showing full compatibility with our previous findings. 

Although experimental data available for orthorhombic phase of SrRuO$_3$ allow to provide a few thoughts on the suitability of applied approximations, the comparison between experiment and theory is not straightforward. It is because of temperature and zero-point phonon effects which in general tend to reduce the elastic moduli. Therefore, one should expect the calculated parameters to be lower at room temperature. For example, the thermal effects can modify the bulk modulus by about 5-15\%, while zero-point phonon contribution can reach 1-5\% for various metals \cite{csonka_2009_21}. Besides, one should also remember that uncertainties in experimental bulk moduli are much greater than in lattice constants and can easily be as large as 10\% (Ref. \onlinecite{staroverov_2004_70}). By taking into account all these considerations, we can conclude that overall our calculated polycrystalline parameters demonstrate a good agreement with available experimental data. Yet, it is not easy to say which of the particular functionals perform best. The general trend observed for the lattice constants of orthorhombic SrRuO$_3$ suggests that it is very likely that revised functionals may yield the closest values to the possible low-temperature experiment, while LDA and PBE may show a slight overestimation and underestimation, respectively.  

\begin{figure}
\includegraphics[scale=0.5]{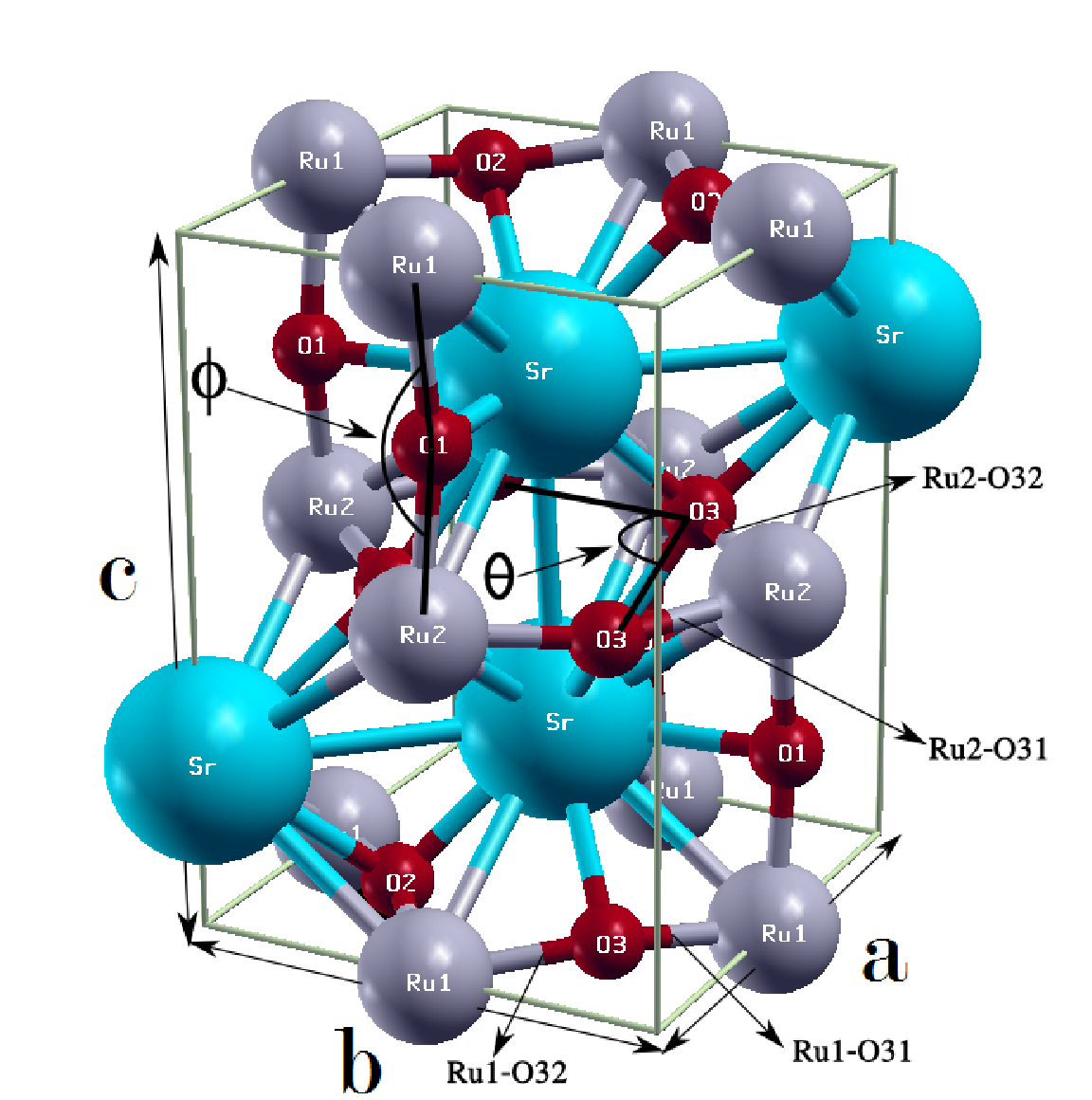}
\caption{\label{fig4}
Schematic representation of the crystalline structure of ground-state orthorhombic ($Pbnm$) SrRuO$_{3}$ under $C_{44}$ related shear deformation. The tilting angle of oxygen octahedra is defined through the relation $\Phi=(180^{\circ}-\phi)/2$, whereas rotation angle is expressed as $\Theta=(90^{\circ}-\theta)/2$. Due to strain-induced structural transformation, non-equivalent rotation angles for Ru1O$_6$ and Ru2O$_6$ exist. The plot was generated with visualization program XCrySDen (Ref. \onlinecite{xcrysden_79}). }
\end{figure}

\subsection{Shear deformation for orthorhombic SrRuO$_3$}

Under $C_{44}$ related shear deformation the orthorhombic structure of SrRuO$_3$ transforms to monoclinic symmetry. This strain-induced symmetry-reduction is accompanied by the change of the number of non-equivalent Ru and O atoms. As can be seen from Fig. \ref{fig4}, non-equivalent Ru atoms, labelled as Ru1 and Ru2, form repetitive layers perpendicular to the c axis, whilst non-equivalent O atoms, named as O1, O2, and O3, are located within the Ru-formed layers (O2 and O3) or between them (O1). The following structural transformation indicates the emergence of two non-equivalent oxygen octahedra, namely, Ru1O$_6$ and Ru2O$_6$. The first one consists of three pairs of O1, O2, and O3 atoms, whereas the second one possesses a pair of O1 and four O3 atoms. The fact that oxygen octahedra are not regular in strain-free SrRuO$_3$ (Ref. \onlinecite{zayak_2006_32}) is also reflected in the structure of Ru1O$_6$ and Ru2O$_6$ implying that a larger set of parameters is required for the complete description of geometry of deformed SrRuO$_3$. However, one should remember that octahedra form a three-dimensional network, thus a change in bond length or angle in one direction necessarily restricts the allowed change in bond length or angle in other directions.  

\begin{figure}
\includegraphics[scale=0.7]{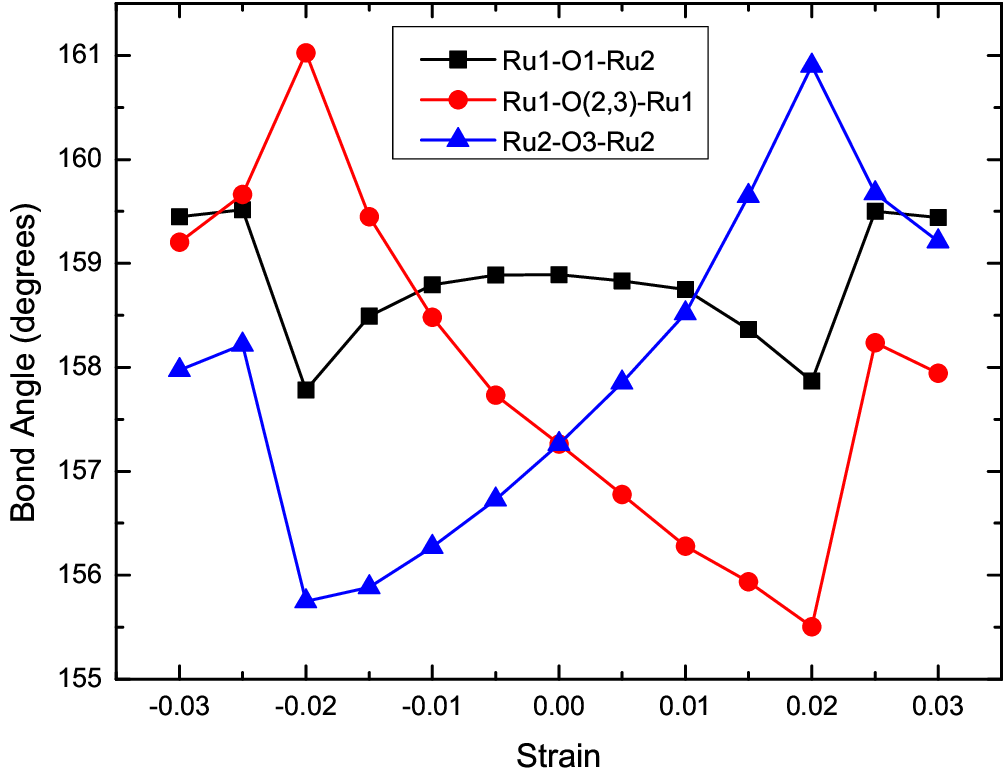}
\caption{\label{fig5}
Ru-O-Ru bond angles of ground-state orthorhombic SrRuO$_3$ as a function of strain under $C_{44}$ related shear deformation. }
\end{figure} 

\begin{figure}
\includegraphics[scale=0.7]{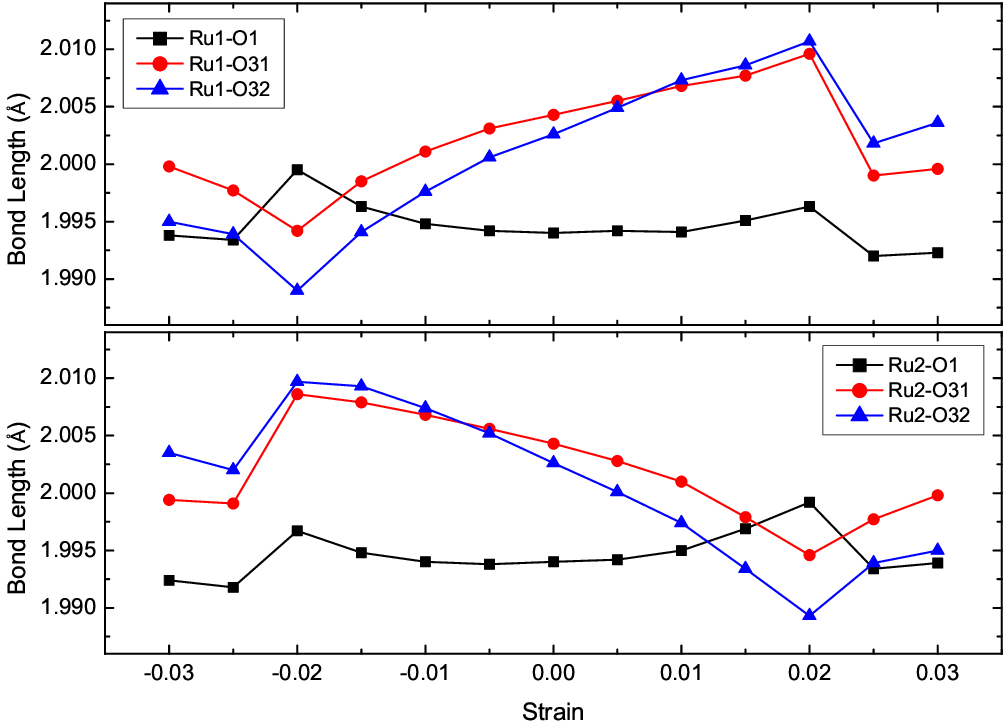}
\caption{\label{fig6}
Ru-O bond lengths of ground-state orthorhombic SrRuO$_3$ as a function of strain under $C_{44}$ related shear deformation. }
\end{figure}

\begin{figure}
\includegraphics[scale=0.6]{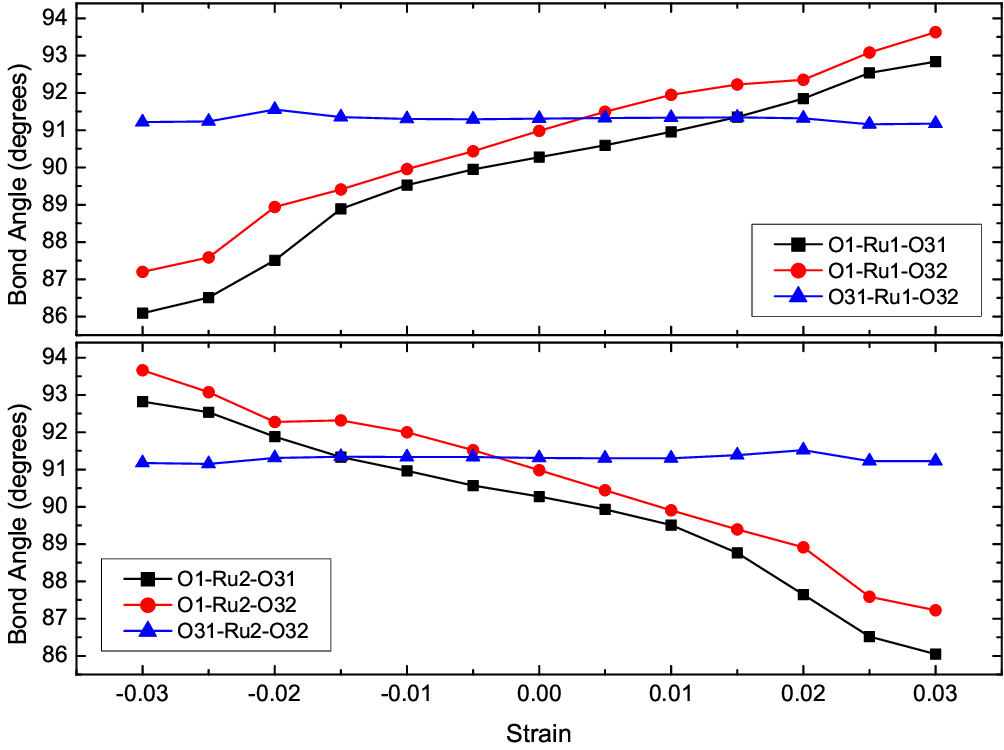}
\caption{\label{fig7}
O-Ru-O bond angles of ground-state orthorhombic SrRuO$_3$ as a function of strain under $C_{44}$ related shear deformation. }
\end{figure}

\begin{figure}
\includegraphics[scale=0.6]{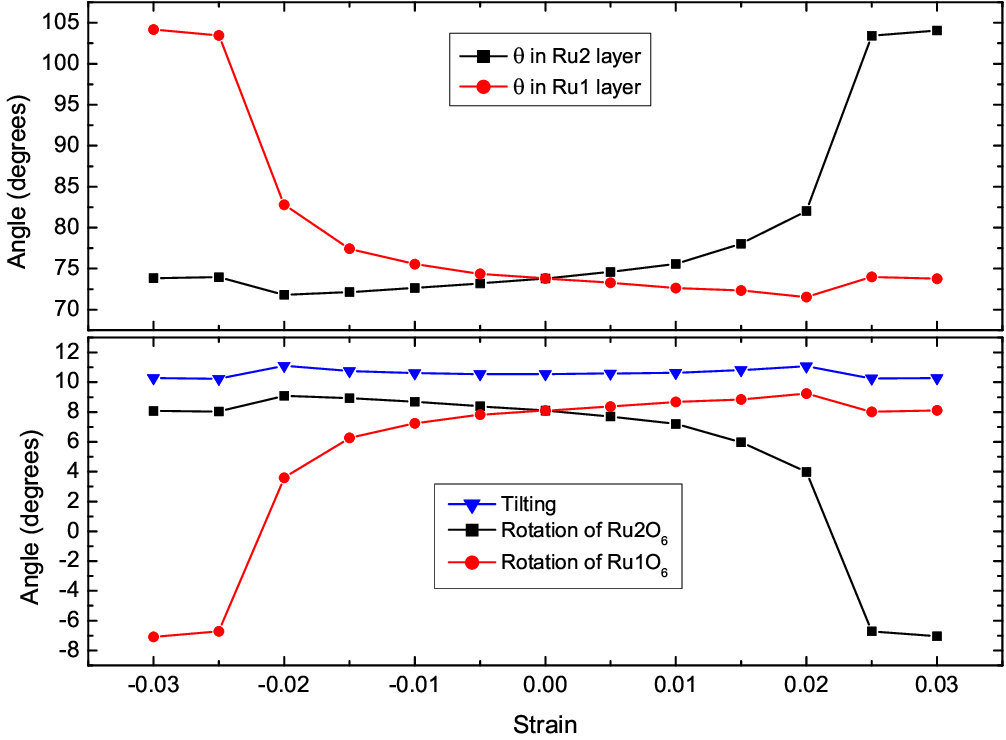}
\caption{\label{fig8}
$\theta$, tilting, and rotation angles of ground-state orthorhombic SrRuO$_3$ as a function of strain under $C_{44}$ related shear deformation.}
\end{figure}

On one hand, structural distortions are highly significant because they have a profound influence on the electronic properties of perovskites. But on the other hand, strain response is in general complex even for ostensibly simple materials meaning that ordinary models in which oxygen octahedra are treated as rigid units or response of octahedral rotations is ignored may not be reliable \cite{rondinelli_2011_71}. For this reason, in Figs. \ref{fig5}, \ref{fig6}, \ref{fig7}, and \ref{fig8} we present a comprehensive set of parameters associated with strain-induced variations in bond lengths and angles between Ru and O atoms. The notation Ru1-O(2,3)-Ru1, introduced in Fig. \ref{fig5}, indicates that bond angles Ru1-O2-Ru1 and Ru1-O3-Ru1 are equal to each other (see Fig. \ref{fig4}). For the sake of simplicity, we did not include bond lengths labeled as Ru1-O21 and Ru1-O22, since they fully correspond to Ru1-O31 and Ru1-O32. The same goes for notation available in Fig. \ref{fig7} which in turn describes the unique bond angles existing in Ru1O$_6$ and Ru2O$_6$ octahedra. The bond lengths with double oxygen indices (e. g., Ru1-O31 and Ru1-O32) can be treated as the shortest distances between an oxygen atom and two equivalent ruthenium atoms located in opposite directions. Data presented in Figs. \ref{fig6} and \ref{fig7} completely define the evolution in internal geometry of oxygen octahedra. One can notice that RuO$_6$ are not regular in unstrained SrRuO$_3$ and their irregularity becomes even more pronounced when magnitude of strains is increased. We would also like to stress the symmetric correspondence between structural parameters under compressive and respective tensile strains. This indicates the independence from strain character; the most important factor is the direction of shear deformation, since $C_{55}$ or $C_{66}$ related deformations (around $b$ or $c$ axis, respectively) do not exhibit deviations from the usual response. Analysis of Figs. \ref{fig5} and \ref{fig6} reveals that strain influence on the Ru-O-Ru bond angles and Ru-O bond lengths cannot be separated because a reduction in bond angle is always compensated for by an increase in appropriate bond lengths, and vice versa. Overall, the complexity of geometric response to strain, seen in Figs. \ref{fig5}-\ref{fig7}, makes it hard to identify the mechanism which could be apparently linked to the extraordinary behavior of SrRuO$_3$. Fortunately, in Fig. \ref{fig7} one can note that unlike the bond angles associated with the apical oxygen O1, O31-Ru-O32 show only negligible variations in magnitude. The following observation is perfectly consistent with data presented in Fig. \ref{fig8}. Here, it is clearly seen that in case of tensile deformation, $\theta$ in Ru2 layer undergoes a huge increase reaching $\sim$21 degrees, while a step between two consecutive strains is only 0.005 - from 0.02 to 0.025. By taking into account that bond angle O31-Ru2-O32 remains almost unaltered, it becomes evident that a significant rotation of Ru2O$_6$ octahedra occurs, with a change in rotation angle $\Theta$ reaching $\sim$11 degrees. In the meantime, variation in rotation of Ru1O$_6$ octahedra as well as tilting stays quite small compared to that of Ru2O$_6$. The opposite trend can be observed as the deformation becomes compressive; in this case, a significant (and approximately equal in magnitude) rotation of Ru1O$_6$ octahedra occurs while rotation of Ru2O$_6$ and tilting show small variations from their initial values. Interestingly, the initial RuO$_6$ octahedral rotation pattern \cite{comment_72} $a^{-}a^{-}c^{+}$ inherent for $Pbnm$ space group is no longer maintained under $C_{44}$ related deformation. It means that adjacent octahedra along $c$ axis begin to rotate in opposite directions. This out-of-phase nature of octahedral rotation is easily recognized from Fig. \ref{fig8}. in which the increase in Ru1O$_6$ rotation angle is followed by the decrease in Ru2O$_6$ rotation angle, and vice versa. 

Although Fig. \ref{fig7} indicates that RuO$_6$ octahedra become less rigid with increasing magnitude of applied strains, one should pay attention that O-Ru-O bond angles demonstrate more-or-less continuous evolution. In contrast, Figs. \ref{fig5}, \ref{fig6}, and \ref{fig8} show an abrupt discontinuity at the critical strain value falling in the range of $\pm$0.02 - $\pm$0.025. This behavior is clearly reflected in deviation from the initial parabola exhibited in Fig. \ref{fig3}. As we have mentioned earlier, the fact that an abrupt change in SrRuO$_3$ behavior is not accompanied by a change in crystallographic space group speaks for the manifestation of isosymmetric phase transition. Although geometrically complex response to strain handicaps the analysis, our observation that tilting angle demonstrates small changes compared to rotation angles and O31-Ru-O32 show tiny variations in magnitude allows us to point out the essential signature of isosymmetric phase transition - vast rotations of oxygen octahedra around $c$ axis. Moreover, the rotation pattern manifests itself in such a way that under tension the Ru2O$_6$ octahedra are much more active in comparison to Ru1O$_6$, whereas under compression the opposite is observed. We would like to notice that this process is simultaneously followed by an increasing deviation in the rigidity of RuO$_6$ thus introducing some uncertainty about the suitability of the deduced scheme. But despite that, strongly pronounced rotations of oxygen octahedra seem to be the most evident mechanism that could be associated with the isosymmetric phase transition in SrRuO$_3$. 

\begin{figure}
\includegraphics[scale=0.7]{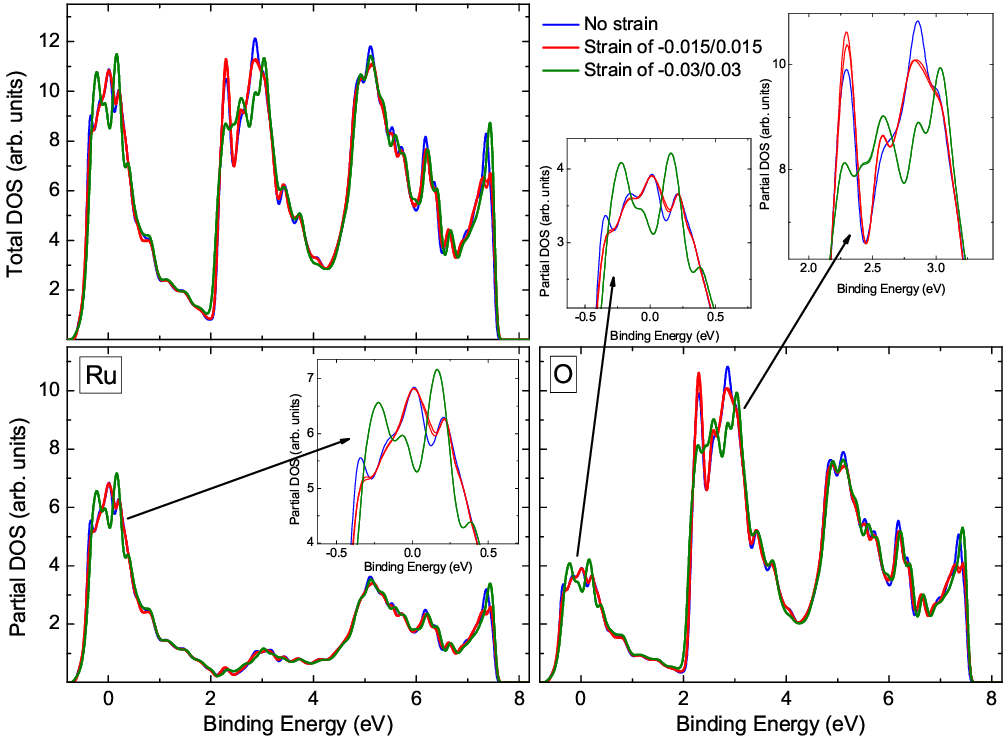}
\caption{\label{fig9}
The total and partial DOS of ground-state orthorhombic SrRuO$_3$ under $C_{44}$ related shear deformation. The Fermi energy is set at zero.}
\end{figure}

In order to find out the influence the isosymmetric phase transition has on the electronic structure, we also include Fig. \ref{fig9} in which the total and partial density of states (DOS) at several values of strain is presented. Notation -0.015/0.015 and -0.03/0.03 indicates that SrRuO$_3$ behavior basically does not depend on whether the applied deformation is compressive or tensile. The insets presented herein reveal an apparent change in the evolution of the spectra, especially in the vicinity of the Fermi energy for Ru and at slightly higher binding energies for O. It is clearly seen that the curves representing DOS of unstrained SrRuO$_3$ and SrRuO$_6$ under -0.015/0.015 deformation exhibit a close resemblance, while a curve of -0.03/0.03 strain substantially changes its profile. Having in mind that the intensity of Ru character at the Fermi level is directly associated with the resistivity of this perovskite \cite{toyota_2005_73}, it becomes obvious that isosymmetric phase transition indeed has a great impact on the electronic and most probably other important properties of SrRuO$_3$. It is interesting to note that isosymmetric phase transition has been recently identified in biaxially strained LaGaO$_3$ (Ref. \onlinecite{coh_2011_74}), LaNiO$_3$ (Ref. \onlinecite{may_2010_75}), and BiFeO$_3$ (Refs. \onlinecite{hatt_2011_76, hatt_2010_77}) films. For these perovskite-structured materials, the isosymmetric phase transition is accompanied by a sharp discontinuity in the magnitude of the tilting and rotation angles which in turn is followed by an abrupt reorientation of the octahedral rotation axis direction. However, no similar trend was observed for biaxially strained orthorhombic SrRuO$_3$ (Ref. \onlinecite{zayak_2006_32}), making our results even more intriguing.

\section{Conclusions}

In this paper, we have studied structural and elastic properties of experimentally observed phases of SrRuO$_3$ by means of standard and revised DFT approximations. The accuracy of the calculated lattice parameters demonstrates that the PBE-like functionals with the parameter $\mu$ restored back to 10/81 can be considered as simple but reliable tool for the investigation of the crystalline structure of SrRuO$_3$. A comparison between the performance of revised functionals reveals a moderately-varying electron density in SrRuO$_3$, whereas a comparison with available theoretical calculations indicates the importance of the properly optimized basis set for Ru. Concerning the elastic properties, no experimental data for single-crystal elastic constants exist, meanwhile the polycrystalline measurements are only available for the low-temperature orthorhombic symmetry. However, our calculations show that polycrystalline SrRuO$_3$ is able to retain its level of compressibility throughout a variety of geometric configurations. In addition, all three phases under investigation here behave in a ductile manner despite tetragonal and orthorhombic SrRuO$_3$ being harder against shear-type distortions. The latter two structures can also be considered as weakly elastically anisotropic, while cubic SrRuO$_3$ exhibits a high degree of anisotropy in shear. Regarding $C_{44}$ related shear deformation, two important issues were clarified. The first one is independent of computational parameters or DFT approaches used and involves mechanical instability of tetragonal SrRuO$_3$ at zero temperature and pressure. In short, $C_{44}$ related shear deformation forces the system to occupy the energetically lower states compared to the strain-free crystal. The mechanism which is responsible for its experimentally observed stability at high temperatures requires a further study. The second issue only quantitatively depends on DFT approximation and involves isosymmetric phase transition of orthorhombic SrRuO$_3$. Briefly, under $C_{44}$ related shear deformation the system displays a distinct deviation from the initial parabolic behavior at the critical strain values of $\pm$0.021 - $\pm$0.024. The mechanism which can be associated with the following discontinuity in behavior is a strongly pronounced out-of-phase rotation of oxygen octahedra around $c$ axis. The isosymmetric phase transition is also accompanied by a variation in electronic structure and very likely other important properties of SrRuO$_3$.     

\begin{acknowledgments}
The authors are thankful for the computational resources provided by the Faculty of Mathematics and Informatics of Vilnius University. \v{S}.M. also gratefully acknowledges the Research Council of Lithuania for the financial support. 
\end{acknowledgments}

\providecommand{\noopsort}[1]{}\providecommand{\singleletter}[1]{#1}%


\begin{thebibliography}{83}
\bibitem [{com({\natexlab{a}})}]{comment_61}%
  \BibitemOpen
  \bibinfo {note} {Pressure is an important
  parameter in the Gibbs free energy, since it can compensate for the
  mechanical instability at zero temperature. The minimization of the Gibbs
  free energy in turn determines the lattice structure of a crystalline
  material in thermal equilibrium at ambient conditions.}\BibitemShut {Stop}%
\end{thebibliography}
\end{document}